\documentclass[sigconf, urlbreakonhyphens=true,authorversion=true]{acmart}

\usepackage[utf8]{inputenc}
\usepackage[nolist,nohyperlinks]{acronym}

\usepackage{xcolor}

\usepackage{listings}
\title{Risk-Based Authentication for OpenStack: A Fully Functional Implementation and Guiding Example}
\subtitle{Data/Toolset paper}

\author{Vincent Unsel}
\orcid{0000-0001-5357-3599}
\affiliation{
    \institution{H-BRS University of Applied Sciences}
    \city{Sankt Augustin}
    \country{Germany}
}
\email{vincent.unsel@h-brs.de}

\author{Stephan Wiefling}
\orcid{0000-0001-7917-6065}
\affiliation{
    \institution{Ruhr University Bochum}
    \city{Bochum}
    \country{Germany}
}
\email{stephan.wiefling@rub.de}

\author{Nils Gruschka}
\orcid{0000-0001-7360-8314}
\affiliation{
    \institution{University of Oslo}
    \city{Oslo}
    \country{Norway}
}
\email{nilsgrus@ifi.uio.no}

\author{Luigi {Lo Iacono}}
\orcid{0000-0002-7863-0622}
\affiliation{
    \institution{H-BRS University of Applied Sciences}
    \city{Sankt Augustin}
    \country{Germany}
}
\email{luigi.lo_iacono@h-brs.de}


 \copyrightyear{2023} 
\acmYear{2023} 
\setcopyright{acmlicensed}\acmConference[CODASPY '23]{Proceedings of the Thirteenth ACM Conference on Data and Application Security and Privacy}{April 24--26, 2023}{Charlotte, NC, USA}
\acmBooktitle{Proceedings of the Thirteenth ACM Conference on Data and Application Security and Privacy (CODASPY '23), April 24--26, 2023, Charlotte, NC, USA}
\acmPrice{15.00}
\acmDOI{10.1145/3577923.3583634}
\acmISBN{979-8-4007-0067-5/23/04}

\begin{acronym}[AAAAAAAA] 

\acro{2FA}{Two-Factor Authentication}
\acro{AP}{Access Point}
\acro{API}{Application Programming Interface}
\acro{AS}{Autonomous System}
\acro{ASGI}{Asynchronous Server Gateway Interface}
\acro{ASN}{Autonomous System Number}
\acro{BASH}{Bourne-again Shell}
\acro{CC}{Country Code}
\acro{CPU}{Central Processing Unit}
\acro{DB}{Database}
\acro{DBMS}{Database Management System}
\acro{DBS}{Database System}
\acro{DOM}{Document Object Model}
\acro{DOS}{Denial-of-Service attack}
\acro{ERD}{Entity-Relation Diagram}
\acro{FV}{Feature Value}
\acro{GUI}{Graphical User Interface}
\acro{HMAC}{Keyed-Hashing for Message Authentication}
\acro{HOTP}{HMAC-Based One-Time Password}
\acro{HTML}{Hypertext Markup Language}
\acro{HTTP}{Hypertext Transfer Protocol}
\acro{HTTPS}{Hypertext Transfer Protocol Secure}
\acro{IaaS}{Infrastructure as a Service}
\acro{IP}{Internet Protocol}
\acro{IPv4}{Internet Protocol Version 4}
\acro{IPv6}{Internet Protocol Version 6}
\acro{ISO}{International Organization for Standardization}
\acro{ISP}{Internet Service Provider}
\acro{JSON}{JavaScript Object Notation}
\acro{LDAP}{Lightweight Directory Access Protocol}
\acro{MAP}{Maximum-A-Posteriori}
\acro{MFA}{Multi-Factor Authentication}
\acro{ML}{Machine Learning}
\acro{NAT}{Network Address Translation}
\acro{NIC}{Network Interface Controller}
\acro{NIST}{National Institute of Standards and Technology}
\acro{NTP}{Network Time Protocol}
\acro{ORM}{Object-Relational Mapping}
\acro{OTP}{One-Time Password}
\acro{RAM}{Read-Only Memory}
\acro{RBA}{Risk-Based Authentication}
\acro{RBAC}{Role-Based Access Control}
\acro{REST}{Representational State Transfer}
\acro{RTT}{Round-Trip-Time}
\acro{SDN}{Software Defined Network}
\acro{SHA}{Secure Hash Algorithm}
\acro{SMTP}{Simple Mail Transfer Protocol}
\acro{SQL}{Structured Query Language}
\acro{SSH}{Secure Shell}
\acro{TLS}{Transport Layer Security}
\acro{UA}{User-Agent}
\acro{UML}{Unified Modeling Language}
\acro{UCS}{Universal Coded Character Set}
\acro{URL}{Uniform Resource Locator}
\acro{UTF-8}{Unicode Transformation Format}
\acro{UUID}{Universally Unique Identifier}
\acro{VCS}{Version Control System}
\acro{VM}{Virtual Machine}
\acro{VPN}{Virtual Private Network}
\acro{WS}{WebSocket}
\acro{WSGI}{Web Server Gateway Interface}

\acro{AP}{Access Point}
\acro{ARP}{Address Resolution Protocol}
\acro{ASCII}{American Standard Code for Information Interchange}
\acro{CLI}{Command Line Interface}
\acro{DHCP}{Dynamic Host Configuration Protocol}
\acro{IEEE}{Institute of Electrical and Electronics Engineers}
\acro{IETF}{Internet Engineering Task Force}
\acro{ISP}{Internet Service Provider}
\acro{MAC}{Media Access Control}
\acro{OSI}{Open Systems Interconnection}
\acro{RFC}{Request for Comments}
\acro{RPC}{Remote Procedure Call}
\acro{SPoF}{Single Point of Failure}
\acro{UDP}{User Datagram Protocol}
\acro{VPN}{Virtual Private Network}

\end{acronym}

\begin{abstract}
Online services have difficulties to replace passwords with more secure user authentication mechanisms, such as \ac{2FA}. This is partly due to the fact that users tend to reject such mechanisms in use cases outside of online banking. Relying on password authentication alone, however, is not an option in light of recent attack patterns such as credential stuffing. %
\ac{RBA} can serve as an interim solution to increase password-based account security until better methods are in place. Unfortunately, RBA is currently used by only a few major online services, even though it is recommended by various standards and has been shown to be effective in scientific studies. This paper contributes to the hypothesis that the low adoption of \ac{RBA} in practice can be due to the complexity of implementing it. We provide an \ac{RBA} implementation for the open source cloud management software OpenStack, which is the first fully functional open source \ac{RBA} implementation based on the Freeman et al. algorithm, along with initial reference tests that can serve as a guiding example and blueprint for developers. \end{abstract}


\begin{document}
\maketitle

\section{Introduction}

Passwords are still the predominant authentication method for most online services~\cite{quermann_state_2018}, despite their long-known weaknesses~\cite{morris_password_1979} and the continuous emergence of new attacks such as credential stuffing and password spraying~\cite{akamai_credential_2019}. To effectively protect their users, online services must use alternative or additional measures to passwords. Other authentication factors using special user-owned devices or physical biometrics, are generally impractical for online services, as they require additional hardware and active user enrollment. This is why they are generally not used in practice~\cite{gaddam_usage_2019}. The composition of two different user authentication factors, usually a password combined with something the user possesses or is, suffers from the same acceptance problems. Apart from few applications such as online banking~\cite{wiefling2020more,reese_usability_2019,dutson_dont_2019}, \acf{2FA}~\cite{petsas_two-factor_2015} is not yet widely accepted by users in practice either~\cite{twitter_account_2021,newman_facebook_2021,milka_anatomy_2018}.

\acf{RBA}~\cite{wiefling2019really} is an online-service-side complement to authentication systems such as password-based authentication that does not require direct user interaction. The user performs the log-in process simply by entering their login credentials (i.e., username and password). In many cases this is sufficient to authenticate to the service. Only when the online service's \ac{RBA} component detects a deviation from the usual log-in behaviour (e.g., different user location), a further authentication factor %
is requested~\cite{wiefling2020more}. This is also true when a correct username-password combination is provided, i.e., an attacker used leaked credentials. Therefore, to keep it with the security principle of ``\emph{good security now}''~\cite{Garfinkel2005DesignPA}, \ac{RBA} can be used as an immediate additional security measure for password-protected user accounts in online services. It enhances the security of such online accounts right away until alternative and more secure authentication methods become established in the mainstream.

\vspace{0.5em}\noindent\textbf{Research Hypothesis.}
Although the use of \ac{RBA} is recommended in the literature~\cite{wiefling2022pump,wiefling2021whats,wiefling2020more} and in national policies~\cite{NIST-SP-800-63-3,biden_jr_executive_2021,national_cyber_security_centre_cloud_2018,australian_australian_2021}, the adoption of \ac{RBA} in practice is still very limited~\cite{wiefling2020more,gavazzi_a_2023}. Recent research found that 78\% of 235 popular online services inside the Tranco 5K~\cite{le_pochat_tranco_2019} still do not use any form of RBA~\cite{gavazzi_a_2023}. As Internet users have an average of 92 to 130 online accounts~\cite{lebras_online_2015}, services of the major Internet companies protect only a fraction of these accounts with RBA. One reason for this low RBA adoption rate could be a lack of guidance on the implementation of \ac{RBA}. Insights from the literature show that effective \ac{RBA} implementations can be very complex~\cite{freeman2016you,wiefling2022pump,bumiller_towards_2022}. Therefore, we expect that many developers cannot estimate the challenges they face when implementing \ac{RBA} in their online services. Furthermore, there are almost no measures available that could help developers with the implementation or testing of a custom \ac{RBA} implementation.

\vspace{0.5em}\noindent\textbf{Contributions.}
To close this gap, we provide the following:
(i) We introduce a conceptual model of \ac{RBA} that provides a generic view on how to implement and integrate RBA in online services. 
(ii) We then instantiate the model as a fully functional open source \ac{RBA} plug-in for the cloud computing software OpenStack. To the best of our knowledge, this is the first open RBA implementation based on the algorithm of Freeman et al. \cite{freeman2016you}.
(iii) Finally, we provide a reference test and reference values based on real world login data that can be used to test RBA implementations based on the Freeman et al. algorithm.

Overall, our work aims to help developers, administrators, and service owners to strengthen password-based authentication by providing guidance on how to implement, integrate, and test RBA for their online services. This should make it easier to put RBA into practice. Our code repository should also serve as a research environment to study how such example implementations help developers to bring security measures into software products.

\section{Risk-Based Authentication (RBA)}\label{sec:risk-based-authentication}

\begin{figure}[t]
    \centering
    \includegraphics[width=0.8\linewidth]{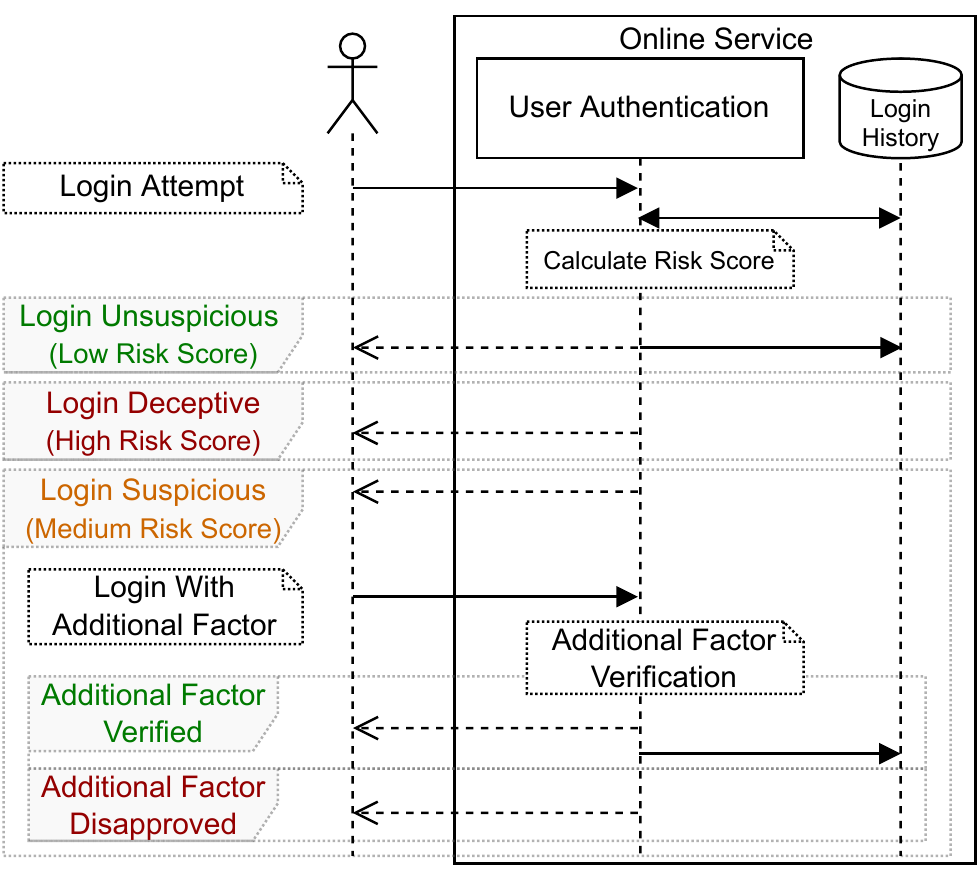}
    \caption{Overview of an RBA system, showing the communication between user and online service in different scenarios (low, medium, and high risk)}
    \label{fig:rba-model}
\end{figure}

Figure \ref{fig:rba-model} shows the common architecture of an \ac{RBA} system~\cite{freeman2016you,wiefling2019really,wiefling2021whats}, typically used in addition to password-based authentication. The system logs contextual features (e.g., network and browser information) for each successful login attempt and compares them to the previously observed feature values. In the case of RBA-enhanced password authentication, this means that access to the service is only granted if the login context is not too different from the previously observed ones. This comparison of the current login context with a recorded history of login contexts is done by calculating a so-called risk score. The score is a number indicating the deviation from the expected values, i.e., the higher the number, the higher the risk of account takeover. Only if this risk score is below a certain threshold, the whole login process behaves like regular password-based authentication from the user's perspective. If the threshold is exceeded, the RBA-based authentication system requires a re-authentication step, i.e., requesting a second authentication factor. Typically, this factor is a verification code sent to a registered email address or phone number~\cite{gavazzi_a_2023,wiefling2019really}.
Only in case this additional factor is successfully verified, the user is authenticated to the service. In case of a very high risk score, e.g., when we are very sure that it is an automated attack, the online service can block access. Although this use case is frequently mentioned in literature, it seems to be uncommon in practice~\cite{gavazzi_a_2023,wiefling2019really}.

The contextual features used in the risk score calculation can be defined by the person configuring the \ac{RBA} system, and can vary from network (e.g., IP address or RTT) and device (e.g., user agent string), to behavioral biometric information (e.g., login time). Related research suggests that network-based (IP address) and device information (user agent string) are sensible features providing high security with high usability in practice~\cite{wiefling2021whats,wiefling2019really,freeman2016you}. When used in combination, these features protected against sophisticated attackers that know the users' passwords, locations, and devices, without legitimate users noticing any differences in the online service's behavior~\cite{wiefling2021whats}. Therefore, we focused on these features in our OpenStack implementation.

To the best of our knowledge, Freeman et al.~\cite{freeman2016you} published the first \ac{RBA} algorithm for computing a risk score for a login attempt. The algorithm was also suspected to be used in some form at the popular online services Google, Amazon, and LinkedIn~\cite{wiefling2021whats}. As it also showed good performance in a practical evaluation and outperformed an algorithm used in the open source single sign-on solution OpenAM~\cite{wiefling2021whats}, we selected this algorithm for our implementation. The algorithm calculates the risk score $S$ for a user $u$ and a given feature set $FV=(FV^1, \ldots, FV^d)$ with $d$ features as:
\begin{equation}\label{eq:rba-without-attack-data}
    S_{u}(FV) = \left( \prod_{k=1}^{d} p(attack | FV^k) \frac{p(FV^k)%
    }{p(FV^k | u, legit)%
    } \right) \frac{p(u | attack)}{p(u | legit)}
\end{equation}

In the formula, we use the probabilities $p(FV^k)$ for the feature value appearing in the global login history of all users, and $p(FV^k | u, legit)$ for the feature value being used by the legitimate user trying to sign in. 
For some features, comparison of occurrence rarely result in an exact match. For example, nowadays, IP addresses are typically assigned dynamically and devices (especially mobile devices) are roaming in different networks. In these cases, when using a unseen feature, $p(FV^k | u, legit)$ would be zero, resulting to an undefined value. To compensate for such frequently changing feature values, sub-feature derivation techniques can be applied to some features as so-called smoothing~\cite{freeman2016you}. Therefore, the IP address feature can be split at three granular levels. Including the \ac{ASN}, to which the address belongs, smooths the unseen feature value to the granularity level of the \ac{ISP}'s \ac{ASN} range \cite{RFC1930-1996}. The third and least granular level is the corresponding country to even compensate frequent rotations of different \acp{AP} \cite{ISO3166-1-2020}. Likewise for smoothing the \ac{UA} string, sub-features are extracted to include browser, operating system and device information into the risk score. These derived sub-features are aggregated in the probability calculation using a linear interpolation~\cite{freeman2016you}, in which the history is divided into all entries containing each sub-feature value to estimate the originated feature value.
Furthermore, $p(u | attack)$ estimates how likely a user is being attacked and $p(u | legit)$ how likely the user is signing in on this online service, i.e., $p(u | legit) =  
\frac{Number\ of\ user\ logins}{Number\ of\ all\ logins}$. 
Some online service might also have additional attack data, e.g., a list of IP addresses that were previously used in attacks. In this case, $p(attack | FV^k)$ estimates how likely the feature value has been previously seen in attacks. Related work indicates, however, that attack data should be used with caution or not at all, as it also can negatively influence the \ac{RBA} system's security and usability properties~\cite{wiefling2022pump}. When not using attack data, this term can be neglected, i.e., $p(attack | FV^k) = 1$~\cite{wiefling2022pump,wiefling2021whats,wiefling2021privacy}.

\section{Related Work}

The Freeman et al. paper~\cite{freeman2016you} can be considered the academic birth of \ac{RBA}. Follow-up academic research provides scientific evidence on useful features~\cite{alaca_device_2016,wiefling2019really,wiefling2021whats,andriamilanto_guess_2021}, user perspective~\cite{doerfler_evaluating_2019,wiefling2020more,wiefling2020evaluation,markert2022howadminsconfigurerba}, suitable \ac{RBA} algorithms~\cite{wiefling2021whats,wiefling2021privacy}, and a large-scale evaluation of effective \ac{RBA} protection based on login data of a real online service~\cite{wiefling2022pump}. To the best of our knowledge, there is no research addressing the reasons for the low adoption of \ac{RBA} in practice or supporting measures for software developers. When analyzing the repositories of relevant open source online software, and identity and access management (IAM) solutions, we found that there are virtually no projects among them that contain RBA capabilities. The IAM software OpenAM\footnote{\url{https://www.openidentityplatform.org/}} is one notable exception. Still, its RBA implementation is very limited in terms of monitored features and risk score calculation. The feature set only includes the IP address, and the current IP address is only compared with those already included in the login history. This approach does not meet the security or usability goals that RBA can otherwise provide~\cite{wiefling2019really,wiefling2021whats,wiefling2022pump}.

Achieving a functional RBA implementation is a very complex task, as it requires a variety of skills and knowledge. To even complicate this, the available scientific literature offers little support for implementation. For instance, the paper by Freeman et al. presents the theory and the algorithm, but many implementation-relevant aspects are left unconsidered and the authors themselves do not provide a reference implementation. The lack of a reference implementation of the Freeman et al. algorithm has since been resolved~\cite{wiefling_basic_2022}, but this is still insufficient as a developer support. The availability of test cases with corresponding test data is often another necessary prerequisite. Therefore, to facilitate the adoption of RBA, we provide a fully functional implementation of RBA in a relevant open source software project. The implementation puts all the necessary pieces together to provide a complete guiding example, including testing capabilities.

\section{OpenStack}

OpenStack\footnote{\url{https://www.openstack.org/}} is an open source software suite for building cloud computing platforms. It offers different Infrastructure-as-a-Service (IaaS) cloud services, primarily virtual machines (VM) and storage services. %
Commercial cloud service providers can use OpenStack to build public cloud platforms and organisations can build private clouds with it.

The OpenStack software is designed as a modular framework composed of different application services. %
As a central component for IaaS systems, the \textit{Nova} compute service manages and hosts virtual instances. Users can administrate Nova through the \textit{Horizon} web dashboard. In addition to Nova, a minimal configuration of OpenStack contains the following modules: the image service \textit{Glance} to discover, register, and retrieve VM images%
; the \textit{Placement} service to enable other services track their own resources; the Software Defined Network (SDN) component \textit{Neutron} to create and attach virtual network infrastructure devices to instances from other modules; and the \textit{Keystone} identity service to enable user authentication and access management.

Relevant service components for integrating RBA functionality are \textit{Horizon} and \textit{Keystone}. The \textit{Horizon} web dashboard allows administrators and users to access and manage cloud computing resources. Authentication requests are delegated to the corresponding identity management service. Horizon offers two possibilities to render the received authentication result in the frontend: When the login credentials were incorrect or the user has to provide an additional additional factor, it can display this information as an error, i.e., a red box in the login form. When access was granted, Horizon redirects to the user's dashboard.

\textit{Keystone} is the access control component for the framework's services. It can identify users and verify their authorization on managed assets. On successful user authentication, Keystone issues a unique login session identifier (session token) to the user. Keystone then constantly verifies whether this token is valid and whether user actions are within their permitted scope.

We chose OpenStack as the example software for providing an RBA implementation. We made this decision because many clouds are built on OpenStack~\cite{miller_brief_2015}. They can therefore immediately benefit from increased account security by installing and using RBA. In addition, OpenStack has a modern microservices-based architecture and built-in extensibility. For this reason, RBA-related source codes are highly decoupled from the other components and can be read and understood with limited knowledge of the overall system. With Python as the underlying programming language, which is also the most popular programming language~\cite{ieee_prog_langs_2022}, the RBA source code should be accessible and comprehensible to most software developers.

\section{OpenStack RBA Extension}
To enable \ac{RBA} in OpenStack, we must extend the Horizon frontend component with RBA-specific login user interfaces and implement the risk score calculation in the Keystone backend component. We leveraged OpenStack's modular microservice architecture and extension interfaces and developed two extension plug-ins for these two components as follows. They work at least with the Wallaby release up to the actual released stable series Zed.

\begin{figure}[t]
	\centering
	\includegraphics[width=\linewidth]{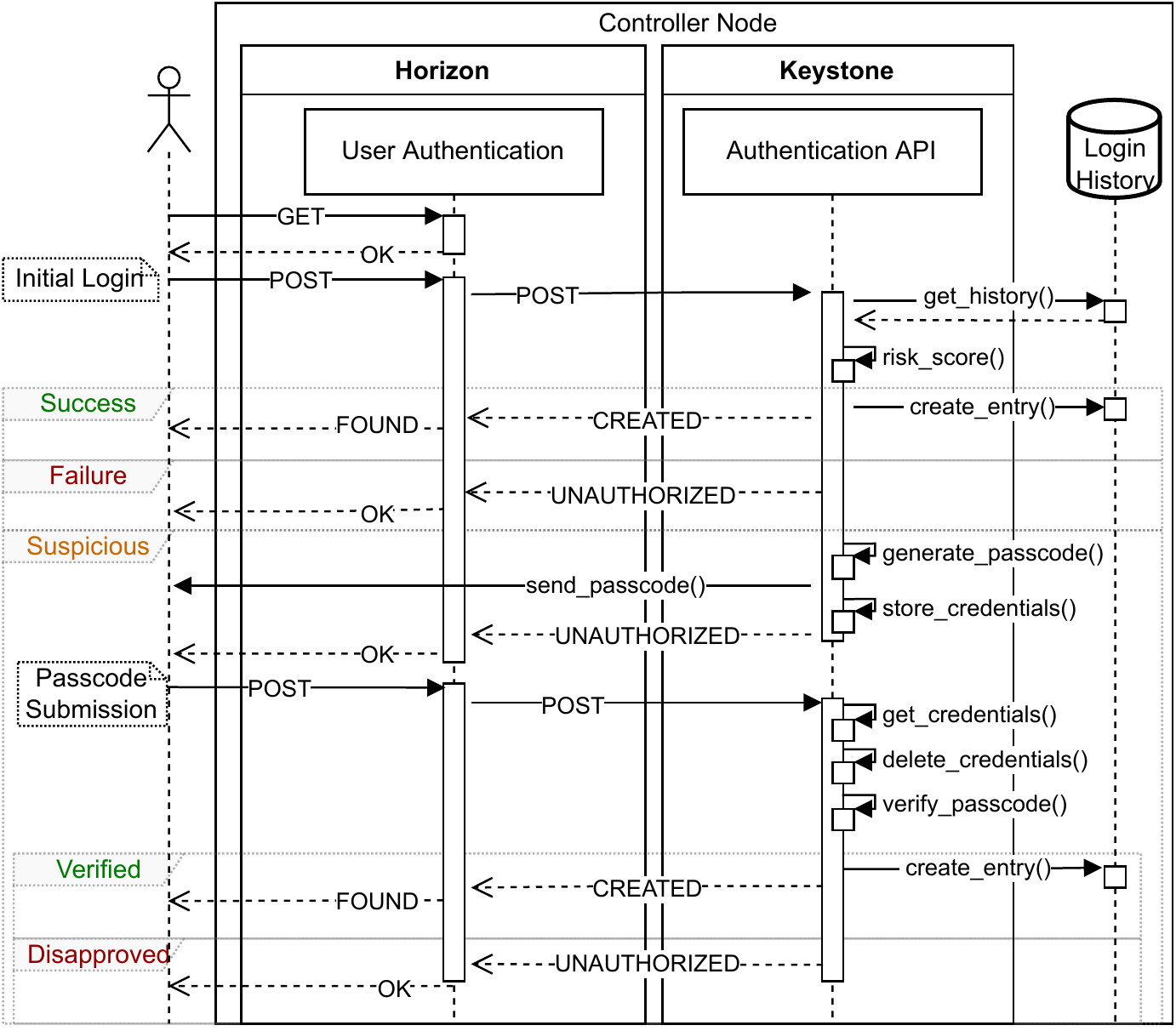} %
	\caption{Authentication Flow in OpenStack using the RBA Extension}
	\label{fig:rba_features}
\end{figure}

\subsection{Horizon RBA Extension (Frontend)}~\label{sec:horizon-rba-extension}
The Horizon \ac{RBA} plug-in\footnote{Available online at \url{https://github.com/das-group/password-rba-horizon}} extends the default password-based authentication by unobtrusively collecting the client's \ac{RBA} features at the login screen. In the current version, the IP address, user agent string, and the \ac{RTT} features are supported by the extension. We chose these features as previous research identified them as the most effective ones to identify users~\cite{wiefling2021whats,wiefling2022pump}. The first two features are obtained directly from the client \ac{HTTP} request. 
The last one required an additional Horizon extension that uses the asynchronous WebSockets protocol for measuring the transmission time of messages from the Horizon service to the client and back. The \ac{RTT} is a strong and privacy-preserving indicator to distinguish legitimate login attempts from those spoofing the ``correct'' location using an \ac{VPN} with an egress point in the same region as the legitimate user~\cite{wiefling2021whats,wiefling2021privacy}. To compensate the network delay jitter, these round-trip measurements are performed five times, and the shortest time measurement is selected as the \ac{RTT} feature in \ac{RBA} authentication~\cite{wiefling2021whats}. 

Figure~\ref{fig:rba_features} shows the RBA-enabled authentication flow in OpenStack. When the user performs a login attempt, Horizon forwards the \ac{RBA} features together with the password to the authentication API of Keystone. The \ac{RBA} evaluation (see Figure~\ref{fig:rba-model}) can have three different results where the Horizon frontend initiates different actions. These are 
(i,~success) the user is authenticated,
(ii,~failure) the authentication has failed, and
(iii,~suspicious) the \ac{RBA} feature verification calculated a suspicious risk score that requires a re-authentication factor to be requested. In our implementation, we use a verification code (passcode) generated by a \ac{HOTP}~\cite{RFC4226-2005} generator, and sent over a separate channel to the user as a re-authentication factor. We chose email based verification with a six digit code in subject line and body, as this variant performed best in a user study~\cite{wiefling2020more}. Also, verification codes are common practice in real-world RBA systems~\cite{gavazzi_a_2023,wiefling2019really}.

\begin{figure}[t]
    \centering
    \includegraphics[width=0.6\linewidth]{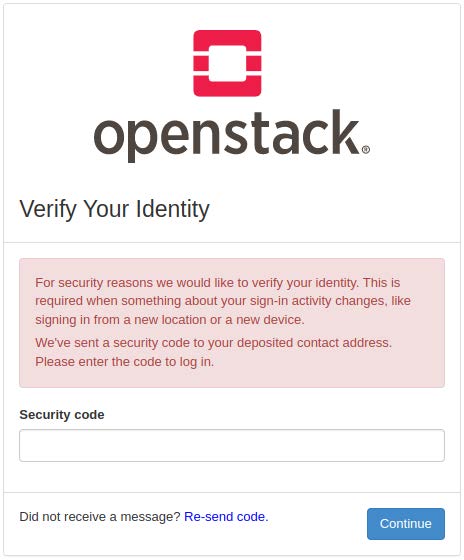}
    \caption{Implemented re-authentication prompt in the Horizon frontend for a suspicious login attempt}
    \label{fig:verification-code-ui}
\end{figure}

In the \textit{suspicious} case, the login dialog requests the verification code that was sent to a previously deposited user contact address, e.g., email address (see Figure~\ref{fig:verification-code-ui}). We based the dialog and verification email on common patterns found in RBA deployments of popular online services, and recommendations of an email verification study~\cite{wiefling2020more, wiefling2020evaluation}. We did this to create an RBA user interface that was tested successfully in multiple usability studies. To be compatible with different contact points in the future (e.g., phone number), we changed the term ``email address'' to ``contact address''. For technical reasons, the user's contact address was not shown in the dialog (see the discussion in Section~\ref{sec:discussion}). The subsequent login attempt using the \ac{RBA} method now contains the verification code instead of features and it needs to match the verification code issued by Keystone to succeed.

\subsection{Keystone RBA Extension (Backend)}
The Keystone \ac{RBA} plug-in\footnote{Available online at \url{https://github.com/das-group/keystone-rba-plugin}} implements the backend \ac{RBA} evaluation based on the features delivered by the Horizon \ac{RBA} frontend, the stored login history, and the Freeman et al. algorithm (see Figure~\ref{fig:rba_features}). 

\vspace{0.5em}\noindent\textbf{Data Collection.}
The \ac{RBA} features received at the beginning of a login attempt are validated and normalized to ensure comparability during further risk score processing. This includes sub-feature extraction to extend the acceptance range for unseen feature values. Finally, the \ac{RTT} feature is normalized by rounding to the nearest ten milliseconds, as suggested by related work~\cite{wiefling2021whats}.

\vspace{0.5em}\noindent\textbf{Risk Score Calculation.}
Our implemented risk score algorithm uses the derived feature values with a linear interpolation probability estimation, as proposed by Freeman et al.~\cite{freeman2016you}, for the calculation of appearance in the user's login history (see Section~\ref{sec:risk-based-authentication}). Although smoothing including sub-features is possible for the probability estimation on the global history, the decision to compare only the originated features was made to limit the needed resources. Otherwise, it would be necessary to either keep a consistent copy of the whole history in-memory or query all entries from the login history database for each login attempt. Both variants seemed not acceptable with a growing user base. For the same reason, a size limit is configurable to even cap the amount of entries for each user. When the limit is exceeded, the oldest entry will be replaced in favor of a new one. 
Instead of narrowing down entries of the whole login history, only the currently attempting user's entries are used and requested from the database. The other estimations can operate on lookup dictionaries or hash tables containing the actual value occurrences in the login history, as suggested by related work~\cite{wiefling2022pump}.

Furthermore, the risk score can optionally include the attack probability for a given IP address. 
We realized this as a third party reputation system, as proposed by Freeman et al.~\cite{freeman2016you}, by using a list containing recently reported malicious network addresses~\cite{fireHOL2022}. Such lists are common for firewall services to automatically block access to protected systems. In our case, we used the IP reputation database from the FireHOL project~\cite{fireHOL2022}, that provides a daily updated collection from several sources. 

\vspace{0.5em}\noindent\textbf{Risk Classification.}
Three different results can be derived from the calculated risk score (see Section~\ref{sec:horizon-rba-extension}). Therefore, two threshold values can be configured, allowing to adjust the occurrence of theses cases. Risk scores below the lower threshold are considered successful authentication and a new entry with the attempt's features is stored in the database.

By exceeding the lower threshold, but still below the upper one, the response indicates Horizon to request an additional verification code. The RBA extension sends the verification code to the registered user via email by default, but this behavior can be changed by setting a different messenger in the configuration (e.g., to send text messages to mobile phone numbers). %

In case even the upper threshold was exceeded, the authentication attempt is rejected. Nevertheless, it is quite common in practice to disable the rejection case~\cite{gavazzi_a_2023,wiefling2019really} and let \ac{RBA} request the re-authentication factor only. Our implementation supports this by setting the rejection threshold to an unreachable high value. 

\vspace{0.5em}\noindent\textbf{Re-Authentication Request.}
In the re-authentication case, Horizon will include the verification code entered by the user in the following authentication request to Keystone. 
If the transmitted verification code could be verified, then a new record will be added to the login history. These previously labeled suspicious feature values will now be taken into account as \textit{``already seen''} on further login attempts.

\subsection{Extending the Feature Set}
We integrated the IP address, user agent string, and \ac{RTT} into the RBA extension, as they proved to be effective to identify users~\cite{wiefling2021whats,wiefling2022pump}. Nevertheless, developers can change the feature set collected by the RBA extension in the code.

To achieve this, they first need to change the list of collected features and their values by the Horizon RBA extension. As Horizon sends these values to Keystone, they also need to adjust this feature set in the Keystone RBA extension. Furthermore, in case of new features beside the three implemented ones, they need to write and connect new validation functions to evaluate the new features.

After that, they need to adjust the two risk score thresholds to values reflecting the risk score values using the new feature set. Adding and removing new features will change the potential range of risk scores, as the amount of multiplications, mostly consisting of probabilities $p$ with $\{p \in \mathbb{R}\ |\ 0 <= p < 1\}$, will change (see Equation~\ref{eq:rba-without-attack-data}). Therefore, we can assume that more features will likely lower the risk scores values. To get an idea of potential values, a reference test can be helpful.

\section{Reference Test}
In order to test a self-developed RBA implementation, some kind of reference test is required. Such a test could provide risk score values obtained by an openly available RBA reference implementation and an openly available login data set. Developers could calculate the risk scores based on the login data set using their own RBA implementation and compare the calculated risk scores to those of the reference implementation. Fortunately, both a reference implementation of the Freeman et al. algorithm~\cite{wiefling_basic_2022} and a \ac{RBA} login data set~\cite{wiefling2022pump} got recently publicly available.

To obtain a reference test from the available resources, one must first determine the reference risk scores. The RBA reference implementation can be used for this purpose. It makes use of the Python pandas~\cite{pandas2020} library that is often used in big data and data science use cases. The reference implementation operates directly on a pandas Data\-Frame object that is initialized with the successful login attempts from the login data set. It provides a test function that calculates risk scores based on all preceding entries at a specified starting point. The result of the test function call contains all risk scores in the interval between the starting entry and the amount of entries to be considered.
Note that the reference implementation allows to calculate risk scores just for slices of the data set. This is a useful feature in early development and testing stages, as it allows to focus the assessments on parts of the data set and reduce high computational costs. The feature also enables scaling the risk score computation across multiple servers and processor cores, e.g., for high performance computing clusters.

\begin{table}[t]
\centering
\caption{Sample of selected risk scores calculated sequentially for successful login attempts of the RBA data set and compared to the risk scores calculated using the Keystone plugin.}
\resizebox{0.8\linewidth}{!}{%
\begin{tabular}{@{}l l l l@{}}
\toprule
\multicolumn{2}{@{}l}{Login Attempt} & & \\
Global & User & Reference Risk Score & Plug-in Risk Score \\

\midrule
64  & 2          &          0.0105382376 &        0.0105382376 \\
67  & 2          &          0.0005499333 &        0.0005499333 \\
89  & 2          &          0.0024253951 &        0.0024253951 \\
... & & \\
10000  & 2       &          0.0184015408 &        0.0184015408 \\
10004  & 2       &          0.0000824648 &        0.0000824648 \\
10008  & 4       &          0.0063803620 &        0.0063803620 \\
10012  & 3       &          0.0002190521 &        0.0002190521 \\
10013  & 9       &          0.0000297278 &        0.0000297278 \\
10014  & 2       &          0.0035949540 &        0.0035949540 \\
... & & & \\
29328 & 4        &          0.0009868327 &        0.0009868327 \\
29331 & 3        &          0.0000687314 &        0.0000687314 \\
29333 & 3        &          0.0000862387 &        0.0000862387 \\
\bottomrule
\end{tabular}%
}
\label{tab:risk_scores}
\end{table}
 
The results do not contain risk scores of login attempts by so far unseen users, as these would be zero anyway. Hence, as soon as a user has logged in more than once, the corresponding risk score is calculated and becomes part of the output. Table~\ref{tab:risk_scores} shows the first calculated risk scores when going through the data set from the beginning. The 64th entry in the data set represents the first recurring login attempt of the same user. Thus, it is the first output of the test function of the reference implementation.

As shown in the table, the implementation of the risk score algorithm in the OpenStack RBA plug-in calculates the same values as the reference implementation. It should be noted that the reference implementation uses the IP address and user agent string and their sub-features ASN, country code, browser name, browser version, operating system name, operating system version and device type from the data set to calculate the risk values. As the validation, normalization, and sub-feature derivation process could result in deviations of feature values from the data set entries used by the implementation to be tested, it is important to circumvent these processes and calculate the risk scores directly on the same feature values contained in the data set.

\section{Discussion}~\label{sec:discussion}
Our guiding example shows that integrating RBA into software projects can be rather complex. We outline some of the issues we faced during implementation in the following.

\vspace{0.5em}\noindent\textbf{Showing the Contact Address in Horizon.}
RBA dialogs typically show the user's email address in censored form~\cite{wiefling2020evaluation,wiefling2020more}. This could help users to determine which email address received the notification while reducing attack surface for attackers, in case they do not know the full email address of the target. Unfortunately, the current code base of Horizon makes it difficult to forward the user's contact address to the dialog. The only way to forward information from the plugin logic to the user interface seems to be via Python exceptions, where the contact address has to be attached to the error message. However, this might entail security and compatibility issues, as we have to parse the contact address correctly while keeping the parsing mechanisms compatible to other variants in the future (e.g., phone numbers). Therefore, to reduce complexity for future implementation versions, we had to decide against showing the contact address in the dialog%
. Future work should enhance OpenStack's code base to allow forwarding variables from the plugin logic to the user interface to solve this problem.

\vspace{0.5em}\noindent\textbf{Complexity in Implementing RBA.}\label{sec:discussion-complexity}
To integrate RBA into their own software projects, developers need to implement the RBA algorithm in the backend and test the algorithm with a data set and compare the output with the reference implementation. When the comparison was successful, they can integrate the data collection in the frontend, and implement the communication between frontend and backend. After that, they can implement the login flow including the re-authentication prompt. Finally, the developers have to set the different thresholds to classify the different risk categories. This can be complex in practice, however.

As derived by the surrounding risk scores in Table~\ref{tab:risk_scores}, a score above 0.003 could be medium risk and 0.018 high risk, as these risk scores are not too common in the sample. We assume that developers will not find this intuitive, as they rather expect numbers like 0.5 for medium risk and 1.0 for high risk. Ways on how to calibrate these scores to understandable values were not described in the Freeman et al. paper. To solve this problem, Wiefling et al.~\cite{wiefling2022pump} suggested a machine learning based algorithm, which returns a risk score threshold for medium risk. This threshold could be used as a baseline to calibrate all risk scores to more readable values%
.

Another source of complexity could be the integration of privacy into RBA. To protect collected features from attacks, several privacy-enhancing methods for the Freeman at al. algorithm were introduced~\cite{wiefling2021privacy,wiefling2022pump}. While most of the methods will not change the risk score, few of them will. Therefore, developers need to check with reference tests whether the privacy enhancements will keep RBA's usability and security properties.

\section{Conclusion and Outlook}
RBA offers good security that service owners should deploy now to increase account protection for their users, until secure and usable alternatives to passwords for online services become a reality. Technical standards, political instruments, and scientific literature support this. What is still lacking, however, is widespread use beyond the few major online services that have been early adopters. Similarly, there is a lack of open implementations that software developers can use as a source of information to tackle the complex task. To address this gap and facilitate the use of RBA in practice, we provided a first fully functional open source RBA implementation based on the Freeman et al. algorithm for the OpenStack cloud management software. On the one hand, the OpenStack RBA plugin can immediately secure many cloud computing platforms and their resources, as well as users. On the other hand, it can serve as a guiding example for developers. To further assist developers in the complex task of implementing RBA for their online services, we provided a way to test the algorithm implementations of Freeman et al. Finally, we want to engage with software developers through our repository who made use of our open RBA implementation in some way. Our goal is to collect qualitative and quantitative empirical data on how useful complete examples are perceived by developers and what other factors play a role in the adoption of security technologies.

Following our OpenStack implementation, we plan to integrate RBA into more open source software projects. This should help to impact a widespread use of RBA in practice to protect more users from attacks like credential stuffing and password spraying.

\bibliographystyle{ACM-Reference-Format}
\bibliography{acmart}
\end{document}